\documentclass[prl,aps,twocolumn,showpacs,superscriptaddress]{revtex4}
\usepackage{bm}
\usepackage{epsfig}

\usepackage{amsmath}
\usepackage{amssymb}
\usepackage{color}

\newcommand{\rlj}[1]{#1}

\newcommand{\ee}{\mathrm{e}}
\newcommand{\ii}{\mathrm{i}}

\newcommand{\CC}{\mathcal{C}}

\newcommand{\DD}{\mathcal{D}}

\begin{document}

\title{Hyperuniformity and phase separation in biased ensembles of trajectories for diffusive systems}
\author{Robert L. Jack}
\author{Ian R. Thompson}
\affiliation{Department of Physics, University of Bath, Bath, BA2 7AY, United Kingdom}
\author{Peter Sollich}
\affiliation{Department of Mathematics, King's College London, Strand, London WC2R 2LS, United Kingdom}

\begin{abstract}
We analyse biased ensembles of trajectories for diffusive systems.  In trajectories biased either by the total activity or the total current,
we use  fluctuating hydrodynamics to
show that these systems exhibit phase transtions into `hyperuniform' states, where large-wavelength density fluctuations are strongly suppressed.
We illustrate this behaviour numerically for a system of hard particles in one dimension and we discuss how it appears in simple exclusion processes. 
We argue that these diffusive systems generically respond very strongly to any non-zero bias, so that
homogeneous states with ``normal'' fluctuations (finite compressibility) exist only when the bias is very weak.
\end{abstract}

\pacs{05.40.-a}

\maketitle

\newcommand{\rhohat}{\hat{\rho}}
\newcommand{\rhobar}{{\overline{\rho}}}

\newcommand{\tobs}{t_{\rm obs}}

\emph{Introduction} -- %
Non-equilibrium systems exhibit diverse collective behaviour and complex emergent phenomena, many of
which have no counterparts at equilibrium.
Even in simple interacting particle systems, one may encounter 
long-ranged correlations~\cite{spohn83}, dissipative ``avalanche'' events with no typical size~\cite{btw}, 
and dynamical phase transitions~\cite{bodineau2004,garrahan2007}.
Theories that capture the universal aspects of these fluctuations are much sought-after, as a route to general
descriptions of non-equilibrium phenomena.
Here, we analyze non-equilibrium ensembles of trajectories~\cite{bodineau2004,lecomte2005,garrahan2007},
defined through constraints on macroscopic observables such as the total current or activity within a given time period.  Phase transitions within these
ensembles occur when such a constraint leads to a qualitative change in macroscopic behaviour~\cite{bodineau2004,garrahan2007,bodineau2008,hedges2009}.  
In diffusive systems~\cite{bertini2001,bertini2005,tkl2007,hurtado2011-pnas,bertini-revs},
we demonstrate transitions into
``hyperuniform'' (HU) states~\cite{torquato2003}, as well as transitions into the macroscopically inhomogeneous
(``phase separated'') states that have previously been found~\cite{bodineau2004,bodineau2008}. 
Hyperuniform states are characterised by anomalously small density fluctuations on large length scales~\cite{torquato2003,florescu2009,zachary2011,berthier2011,man2013,chicken2014,levine-arxiv}; they
have been identified in jammed particle packings~\cite{berthier2011,zachary2011} and in biological systems~\cite{chicken2014}. 
These systems are highly optimised in response to a global constraint (mechanical stability in jamming,
optimal fitness in biology).  The constrained dynamical ensembles that we consider in this study are also optimised: they are the maximally probable
states consistent with the constraint.  Our results (i)~provide further evidence that hyperuniformity is generic, by demonstrating that it occurs 
in a new set of optimised non-equilibrium ensembles,  and (ii)~resolve
the physical interpretation of some phase transitions that have been previously discovered in diffusive systems~\cite{appert2008,bodineau2008}.

\newcommand{\fdm}{BHPM}

\emph{Models} -- %
We study biased ensembles of trajectories both computationally and analytically.  For computational studies, we consider a one-dimensional model of $N$
diffusing hard particles in a periodic 
 box of size $L$, with each particle having size $l_0=1$.  This
\rlj{Brownian hard-particle model (\fdm)}  evolves by
Langevin dynamics: the position $x_i$ of particle $i$ obeys $\partial_t x_i = -\beta \nabla_i U + \eta_i$
where the $\eta_i$ are independent white noises, $U$ is the potential energy, and $\beta$ the inverse temperature.  
The  diffusion constant of an isolated free particle is $D_{0}$.  
We use a Monte Carlo (MC) dynamical scheme to simulate this system.  
Full system details are given in Appendix A. 

We also consider lattice-based exclusion models where $N$ particles are distributed over $L$ lattice sites, again with periodic boundaries.  At most one particle may occupy any lattice site.  In the partially asymmetric simple exclusion process (PASEP), particles hop left with rate $\ell$ and right with rate $r$, provided their destination site is empty.  The symmetric simple exclusion process (SSEP) is the case $\ell=r=1$.   The steady states of the \fdm\ and the PASEP have no correlations between particles beyond hard-core exclusion. (Unlike the SSEP and \fdm, the PASEP does not obey detailed balance, but for periodic boundaries, it may still be shown that site occupancies are uncorrelated in the steady state.)

\emph{Biased ensembles of trajectories} -- %
Let $K=K[x(t)]$ be a measure of dynamical activity in a trajectory $x(t)$.  For exclusion processes, $K$ is the total number 
of particle hops in a trajectory.  For the \fdm, we follow~\cite{hedges2009}: we choose a coarse-graining time $\tau_0$ and focus on trajectories of length $\tobs = M\tau_0$, defining $K =   \sum_{j=1}^M \sum_{i=1}^N |\hat{x}_i(t_j) - \hat{x}_i(t_{j-1})|^2$ with $t_j = j\tau_0$.  The position $\hat{x}$ is defined by subtracting the centre-of-mass motion (see Appendix A) 
which helps to minimize finite-size effects.  
\rlj{We take $\tau_0=\ell_0^2/(2D_0)$, in which time an isolated particle diffuses a distance comparable with its size.  
We fix the units of time by setting $\tau_0=1$.}

To investigate trajectories that are constrained to non-typical values of $K$, we
define a biased ensemble of trajectories~\cite{spohn1999,lecomte2005,garrahan2007}, via a formula for the average of an observable $O$: 
\begin{equation}
\langle O \rangle_s = \ee^{-\psi_K(s)L^d\tobs} \langle O \ee^{-sK} \rangle_0 .
\label{equ:s-ens}
\end{equation}
Here $\langle \cdot \rangle_0$ represents an average in the (unbiased) steady state of the model, $\langle O \rangle_s$ is an average within the biased ensemble, and $\psi_K(s)=\log \langle \ee^{-sK} \rangle_0/(L^d\tobs)$ is a `dynamical free energy'.  For sufficiently large $\tobs$, averages in the biased ensemble are equal to averages over trajectories in which the activity $K$ is constrained~\cite{touchette2013}. 

\begin{figure}
\includegraphics[width=8.4cm]{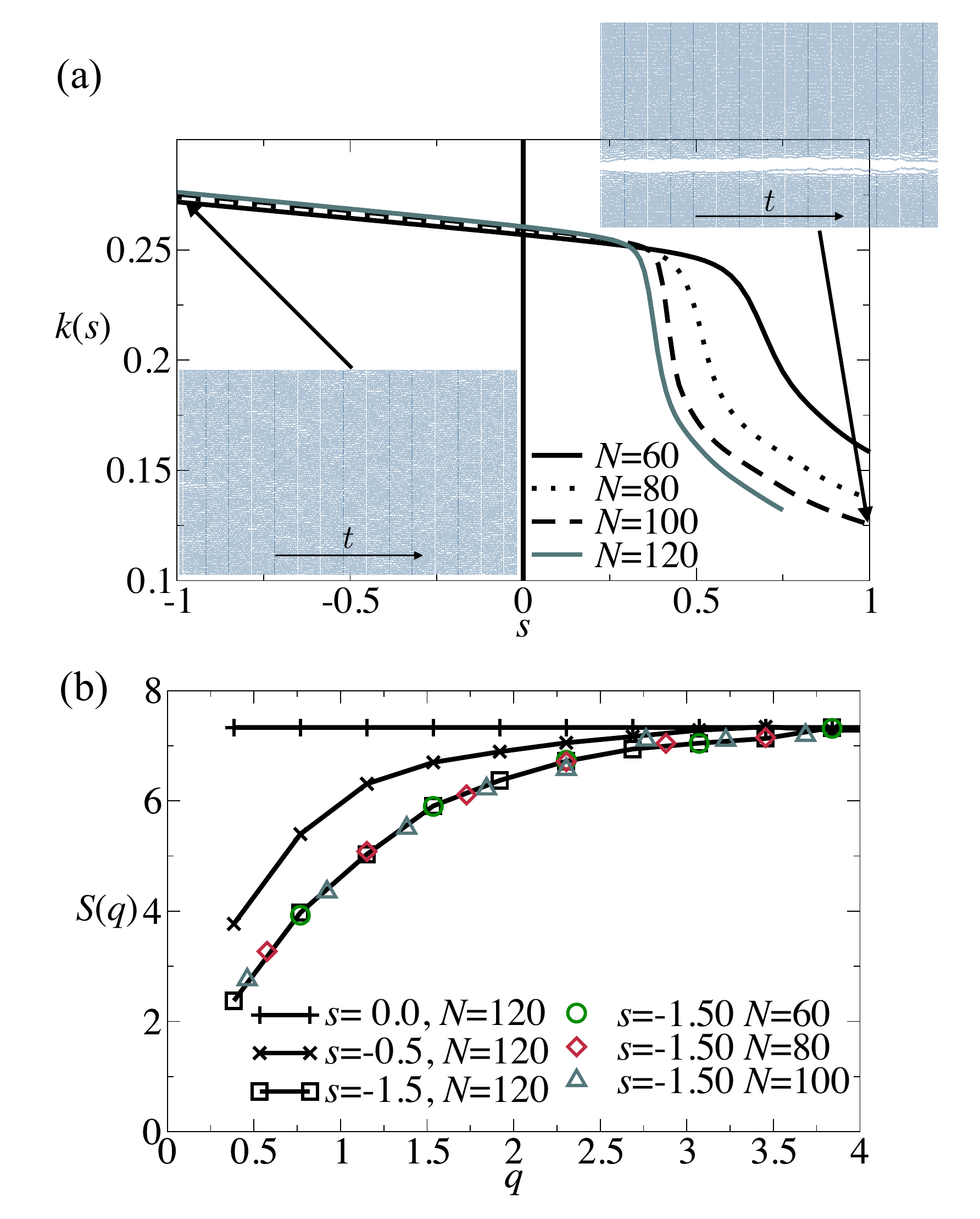}
\caption{Numerical results for the \fdm.  
(a)~Mean activity $k(s)$ in biased ensembles at $\rho=0.88$, with representative trajectories illustrated in space-time (time $t$ runs from left to right).  For $s>0$, phase separation occurs, accompanied by a jump in $k(s)$, while for $s<0$ the system is homogeneous.  (b)~Structure factor $S(q)$ in biased ensembles.  For $s<0$, small-$q$ density fluctuations are strongly suppressed. 
}
\label{fig:fdm}
\end{figure}

\emph{Numerical results for the \fdm} -- %
Fig.~\ref{fig:fdm} 
has results for the \fdm, calculated using transition path sampling~\cite{tps,hedges2009}.  
Fig.~\ref{fig:fdm}(a) shows the mean activity $k(s)=(L^d\tobs)^{-1}\langle K \rangle_s$.  For $s>0$ there is a first-order transition into a phase separated state~\cite{bodineau2004,bodineau2008,lecomte2012}.  For $s<0$, the activity appears to depend smoothly on $s$, but the system develops strong long-ranged correlations.
These are measured by the structure factor $S(q) =  \frac{1}{L^d} \langle \delta\rho_{q}(t) \delta\rho_{-q}(t) \rangle$ where $\delta\rho_q = \int\mathrm{d}r \delta\rho(r) \ee^{-iq\cdot r}$ and $\delta\rho(r)=\rho(r)-\rhobar$, with $\rhobar$ the mean density.
To see the relevant behavior most clearly, we transform co-ordinates so that the particles are treated as point-like (see Appendix A): 
defining $L_0=L-Nl_0$, the equilibrium ($s=0$) ensemble then has $S(q)=N/L_0$, independent of $q$.  
Fig.~\ref{fig:fdm}(b) shows that for $s<0$ and small $q$, the structure factor deviates strongly from this equilibrium value.  
The signature of a hyperuniform state is that $S(q)\sim q$ at small-$q$~\cite{torquato2003}: density fluctuations on large length scales are strongly suppressed.  
This means that particle positions necessarily have long-ranged correlations [otherwise, self-averaging of the density within large regions of the system implies $\lim_{q\to0} S(q)>0$].
Analysis of the small-$q$ behaviour in numerical simulations is limited by the system size, 
but the results for $s<0$ are consistent
with hyperuniformity.

\emph{Fluctuating hydrodynamics} -- %
The \fdm\ is representative of a general class of diffusive systems, which may be described by ``fluctuating hydrodynamics''~\cite{spohn83,eyink1990,bertini-revs}.
Within this theory, the time-evolution of the density $\rho(r,t)$ on large length and time scales can be approximated by a Langevin equation 
\begin{equation}
\partial_t \rho(r,t) = \nabla\cdot D[\nabla\rho(r,t)-a] + \nabla\cdot[\sqrt{\sigma}\eta(r,t)] ,
\label{equ:fh-langevin}
\end{equation}
 where $\eta$ is a white noise, $D=D(\rho(r,t))$ and $\sigma=\sigma(\rho(r,t))$ are local measures of diffusivity and mobility, and $a$ is an asymmetric driving force.  Details of the relationships between fluctuating hydrodynamics and the \fdm, SSEP and PASEP are given in Appendix B. 
 Note that the fluctuating hydrodynamic theory 
 is valid in all dimensions, not just $d=1$.

\newcommand{\kkap}{\kappa}

\emph{Hyperuniformity within fluctuating hydrodynamics} -- %
Consider a system described by (\ref{equ:fh-langevin}) with $a=0$, and introduce a bias to larger-than-average activity, $s<0$. 
Averages within the biased ensemble are given by path-integral expressions: $\langle O \rangle_s = \ee^{-\psi_K(s)L^d\tobs} \int \DD\rho \DD\rhohat\, O[\rho] \ee^{-\int\mathrm{d}r\mathrm{d}t {\cal L} }$, where
$\rhohat$ is a (real-valued) response field, and
\begin{equation}
{\cal L} = \ii\rhohat [ \partial_t \rho - \nabla\cdot (D\nabla\rho) ] + \tfrac12 \sigma (\nabla\rhohat)^2 + s\kkap,
\label{equ:fh-lagrange}
\end{equation}
in which \rlj{$\kkap=\kkap(\rho)$} is the (density-dependent) local activity of the system.  
We assume $\kkap''(\rho)\leq0$, which certainly holds for exclusion processes and may be expected to
hold for generic particle systems;
 analysing the case with $\kkap''(\rho)>0$ is also straightforward~\cite{kmp,lecomte2010,hurtado-rev}. The behavior of $\kkap(\rho)$ for the \fdm\ is shown in Appendix A.

Analysis of hydrodynamic behaviour requires a suitable rescaling of space and time co-ordinates. To avoid 
cumbersome notation we defer this procedure to Appendix B 
and quote our results in terms of the bare (unrescaled) parameters.  Note, however,
 that these results apply only in the hydrodynamic limit.
For $s\leq0$, 
the path integral is dominated by trajectories where $\rho(r,t)\approx\rhobar$ and $\rhohat\approx0$ so we write $\rho(r,t)=\rhobar + \delta\rho(r,t)$ and expand to quadratic order in $\delta\rho$ and $\rhohat$.  The result is
\begin{equation}
{\cal L} \approx \ii\rhohat(\partial_t - \nabla\cdot D_0\nabla)\delta\rho + \tfrac12 \sigma_0(\nabla\rhohat)^2 + s \kkap_0 
+ \tfrac12 s\kkap''_0\delta\rho^2 ,
\label{equ:lag-quad}
\end{equation}
where we write $\kkap(\rho) = \kkap_0 + \kkap_0' \delta \rho + \kkap_0'' \delta\rho^2/2 + \dots$, with $\kkap_0 = \kkap(\rhobar)$, $\kkap_0'=(d/d\rho)\kkap(\rhobar)$ 
etc, and similarly for $D(\rho)$ and $\sigma(\rho)$.

The structure factor may then be evaluated 
(see~\cite[Equ.~(58)]{bodineau2008} and also Appendix B, 
yielding 
\rlj{
\begin{equation}
S(q) = \frac{ \sigma_0 q^2 }{ 2 \sqrt{(D_0 q^2)^2 + s q^2 \sigma_0 \kkap''_0}}.
\label{equ:fh-Sq}
\end{equation}
}%
We again emphasise that this result is valid only for small $q\ll 1$, and that  $s,\kkap''_0<0$, by assumption.

Equ.~(\ref{equ:fh-Sq}) demonstrates a singular response to the field $s$.  
For $s=0$ and $q\to0$, the structure factor approaches a non-zero constant \rlj{$\sigma_0/(2D_0)$}, as expected in an equilibrium state with a finite compressibility.  However, for any $s<0$, the large scale behaviour changes qualitatively: \rlj{$S(q)=(q/2)\sqrt{\sigma_0/(s\kkap''_0)}+O(q^2)$}.  The numerical results of Fig.~\ref{fig:fdm}(b) are consistent with this theoretical prediction.
Note that hyperuniformity is a large length scale phenomenon: the non-trivial behaviour in $S(q)$ appears only for small $q\lesssim \sqrt{s\sigma_0 \kkap_0''}/D_0$.

We also calculate the mean activity $k(s) = \langle\kkap\rangle_s \approx \kkap_0 + (\kkap_0''/2)\langle\delta\rho(r,t)^2\rangle_s$.   Writing $\langle\delta\rho(r,t)^2\rangle = \int\frac{\mathrm{d}q}{(2\pi)^d} S(q)$, we see that
the suppression of $S(q)$ at small $q$ acts to increase $k(s)$ [recall $\kkap_0''<0$].
Taking $s<0$ and $d=1$, we obtain (see~\cite{appert2008} and also Appendix B):
\rlj{
\begin{equation}
k(s) - k(s=0) \approx \sqrt{s \kkap_0'' \sigma_0 } \cdot \frac{|\kkap_0''| \sigma_0}{ 4\pi D_0^2 } ,  
\label{equ:k-sqrt-s}
\end{equation}
}%
which is valid to leading order in $|s|$. Since $k(s) = -\psi'_K(s)$ where $\psi_K$ is the dynamical free energy, we identify this non-analytic behaviour in $k(s)$ with a second order dynamical
phase transition.  This singular behavior has been noted before~\cite{appert2008}, 
but its link with hyperuniformity has not.  In $d>1$, the suppression of $S(q)$ at small wavevectors leads to a singular contribution $K(s) - K(s=0) \sim (-s)^{d/2}$, with logarithmic corrections
if $d$ is even (see Appendix B). 
As illustrated in Fig.~\ref{fig:fdm},  biasing to lower-than-average activity by choosing $s>0$ instead leads to phase separation~\cite{bodineau2004,bodineau2008,lecomte2012,hurtado-rev}.  

\emph{Heuristic explanations for HU states} -- %
The origin of hyperuniformity in biased diffusive systems 
is the diverging hydrodynamic time scale associated with large-scale density fluctuations.  
To see this, consider linear response to the field $s$.  Within a biased ensemble of trajectories, the probability of finding the system in a configuration $\CC$ is
$p_\CC(s) = p_\CC(0) [1 - 2 s \int\mathrm{d}r\mathrm{d}t \langle \delta \kkap(r,t) \rangle_{\CC} + O(s^2)]$ where $ \langle \delta \kkap(r,t) \rangle_{\CC}$ is a ``propensity''~\cite{propensity}, which is obtained by averaging the activity over trajectories that start in $\CC$ at $t=0$, and comparing with typical equilibrium trajectories~\cite{garrahan2009,jack2014-east}. 

If  $\CC$ has an unusual density fluctuation at a small wavevector $q\approx 1/R$, expanding $\delta \kkap$ to quadratic order in $\delta\rho$ gives $\int \mathrm{d}r \delta \kkap(r,t)  \approx \frac{\kkap_0''}{2L^d} [ |\rho_q(t)|^2 - \langle  |\rho_q(t)|^2 \rangle_0 ]$.  Diffusive scaling therefore indicates that $\langle \delta \kkap(r,t)\rangle_{\CC} \approx \frac12 \kkap_0'' [|A_\CC|^2 - S_0(q)] \ee^{-t/\tau_R}$, where $\tau_R=R^2/D_0$ is a relaxation time, $A_\CC=\rho_q(0)/L^{d/2}$ is the amplitude of the density fluctuation and $S_0(q)$ the structure
factor of the unbiased state.  
In hyperuniform states we expect $|A_\CC|^2 \ll S(q)$ yielding $(d/ds) \left. p_\CC(s)\right|_{s=0} \propto \kkap_0''S_0(q) \tau_R$: the time scale $\tau_R \sim R^2$ 
diverges for large $R$ and $\kkap''<0$, so these $p_\CC$ are strongly enhanced for $s<0$.
On the other hand, phase-separated configurations have $R\approx L$ and  $A \sim L^{d/2}$ so $(d/ds)  \left. p_\CC(s)\right|_{s=0} \propto -\kkap_0'' L^{d}\tau_L$: these configurations have strong (divergent) enhancement  for $s>0$ and as $L\to\infty$.  Hence, this perturbative analysis
reveals an instability of the small-$q$ modes to changes in $s$: we argue that this is the origin of the HU and phase-separated states when $s\neq 0$.  The diverging diffusive time scale $R^2/D_0$ is central to this analysis, similar to other cases where diverging time scales lead to phase transitions in biased ensembles~\cite{jack2010rom,garrahan2009}.

\begin{figure}
\includegraphics[width=8cm]{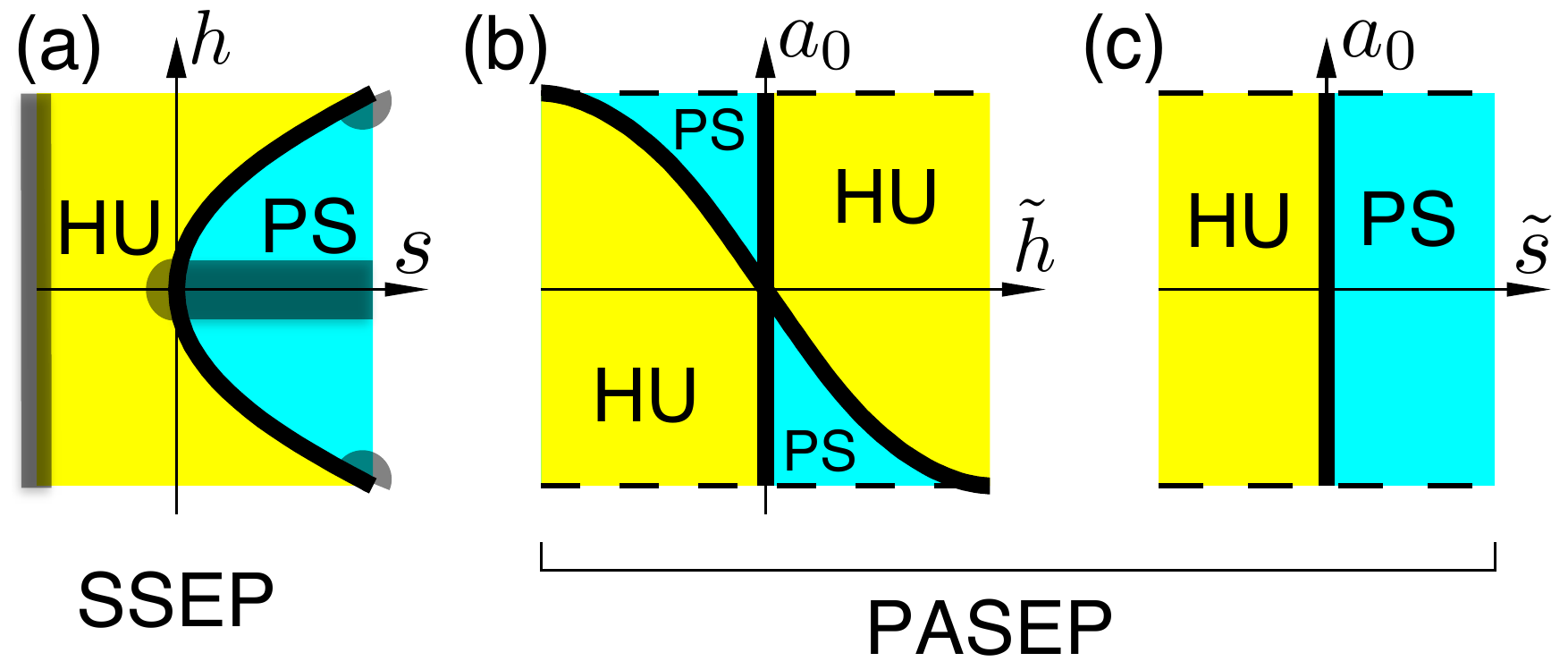}
\caption{Proposed dynamic phase diagrams for biased exclusion processes. HU: hyperuniform states; PS: phase separation. (a)~SSEP, jointly biased by both activity and current. On the heavy black line, the system has normal fluctuations. In the shaded regions, the indicated behavior can be shown analytically. (b)~PASEP with hopping asymmetry $a_0$, biased by the current.  
Normal fluctuations occur for zero bias ($\tilde{h}=0$) and on the line $a_0=-\tanh(\tilde{h}/2)$.
(c)~PASEP, biased by the activity: normal fluctuations occur only for zero bias, $\tilde{s}=0$.}
\label{fig:phase}
\end{figure}

\emph{Biased ensembles based on the total current} -- %
So far, we have considered ensembles of trajectories biased according to their activity $K$.
In fact, HU states also appear in ensembles of trajectories where the total current is biased.  We define the total current $J$ as the sum of all (directed) particle displacements
in a trajectory.  (For exclusion processes, this is the difference between the numbers of right- and left- hops.)  For generality, we consider jointly-biased ensembles where the activity is biased
by a field $s$ and the current $J$ is biased by a field $h$.  The analogue of (\ref{equ:s-ens}) is $\langle O \rangle_{s,h} = \ee^{-\psi_{\rm KJ}L^d\tobs} \langle O \ee^{hJ-sK} \rangle_0$ (see Appendix B). 
Within fluctuating hydrodynamics and assuming a time-reversal symmetric unbiased state ($a=0$)
one finds 
that the response to the bias depends only on the quantity $B=s\kkap_0'' - \frac12 h^2\sigma''_0$ (see Appendix B).  
For $B=0$ then $S(q)$ has a finite (non-zero) limit as $q\to0$; for $B>0$ the system is HU while for $B<0$ one has phase separation.  The resulting dynamical phase diagram is shown in Fig.~\ref{fig:phase}: the fluctuating hydrodynamic analysis holds only for $s,h\ll 1$, but in the absence of additional phase transitions one expects the same structure to hold throughout the $(s,h)$-plane.  We discuss this conjecture below, using results from exclusion processes.  We also note in passing that the centre of mass in the \fdm\ undergoes free diffusion so that the distribution of the current $J$ is trivial in that case: 
in the fluctuating hydrodynamic theory, this means that $\sigma\propto\rho$, so $\sigma''_0=0$.

The condition $B=0$ [solid line in Fig.~\ref{fig:phase}(a)] recovers a homogeneous state with $S(q\to0)>0$: we use the term ``normal fluctuations" for this case, in contrast to hyperuniform or phase-separated states.
In fact, biased ensembles with $B=0$ are identical to (unbiased) steady states of models in which time-reversal symmetry is broken ($a\neq0$). This may be verified directly from the hydrodynamic Lagrangian (\ref{equ:fh-lagrange}) but a clearer interpretation of this result can be obtained by analyzing exclusion processes, as we now discuss.

\newcommand{\sig}{\sigma}

\emph{Mappings between biased ensembles for exclusion processes} -- %
We analyse exclusion processes via operator representations of their master equations~\cite{stinch-review}.
Starting from the master equation for the SSEP, we write a biased generator~\cite{spohn1999,lecomte2005} that describes the jointly biased ensemble.  
This operator has a representation in terms of Pauli spin matrices~\cite{tkl2008} (see Appendix C)
\begin{equation}
W_{\rm S}(s,h) = \sum_i \ee^{h-s}\sig^-_{i}\sig^+_{i+1} + \ee^{-h-s}\sig^-_{i}\sig^+_{i-1}  - 2 n_i (1 - n_{i+1} ),
\label{equ:Wsh}
\end{equation}
where $n_i = \sigma^+_i \sigma^-_i$.
If we consider instead a PASEP biased by its current, the relevant operator is
$W_{\rm A}(\ell,r,\tilde h)=\sum_i r\ee^{\tilde h}\sig^-_{i}\sig^+_{i+1} + \ell\ee^{-\tilde h}\sig^-_{i}\sig^+_{i-1}  - (r+\ell) n_i(1-n_{i+1}) $,
where $\ell,r$ are hopping rates and $\tilde{h}$ the biasing field. 
For appropriate parameters (including always the normalization $r+\ell=2$), we may have $W_{\rm S} = W_{\rm A}$, which means that the trajectories of the two biased
ensembles are identical.
Defining the hopping asymmetry $a_0=(r-\ell)/(r+\ell)$,
equality between $W_{\rm S}$ and $W_{\rm A}$ requires  $a_0=\tanh h$ and $\ee^{s} = \cosh(h-\tilde h)$. 
Any jointly-biased  SSEP with $s>0$ leads to two solutions for $\tilde h$, which correspond to two possible current-biased PASEPs.  These two possibilities are related by a 
Gallavotti-Cohen symmetry~\cite{gallavotti-cohen,spohn1999}. In the specific case $\tanh(\tilde{h}/2)=-a_0$ [so that
$\ee^{\tilde h} = \ell/r$], one has $W_{\rm A}(\ell,r,\tilde h) = W_{\rm A}(r,\ell,0)$: the Gallavotti-Cohen symmetry relates biased and unbiased PASEP states, both of which have ``normal'' fluctuations. 

The SSEP with $\cosh h=\ee^{s}$ is another special case, because it leads to $\tilde h=0$: the steady state of a jointly-biased SSEP corresponds exactly to that of an unbiased PASEP. 
This is the microscopic interpretation of the condition $B=0$ in the fluctuating hydrodynamic analysis [for small $h,s$, we obtain
$s=h^2/2$ which is consistent with $B=0$, because $\sigma(\rho)=\kkap(\rho)$ for the SSEP].
The unbiased steady state of the PASEP has normal fluctuations'', so we conclude that fluctuations are also normal along the line $\cosh h=\ee^{s}$ in the dynamical phase diagram of the jointly-biased SSEP [solid line in Fig.~\ref{fig:phase}(a)].  

There are a family of mappings between biased SSEPs and PASEPs (see Appendix C): 
for example, an activity-biased PASEP may also be mapped to a jointly-biased SSEP.  The resulting situation is shown 
in Fig.~\ref{fig:phase}(b,c) where we show the PASEP dynamical phase diagrams that correspond to the 
(conjectured) phase diagram in Fig.~\ref{fig:phase}(a).
%
The hypothesis is that all points in Fig.~\ref{fig:phase}(a) are either phase separated (PS) or hyperuniform (HU), except for the normal line $\cosh(h) =  \ee^s$.
We provide several arguments in support of this picture.  The fluctuating hydrodynamic analysis establishes these results in the small bias regime, $|h|,|s|\ll1$, since the condition $\cosh h=\ee^{s}$ then reduces to the case $B=0$ discussed above.
The question is therefore whether some other phase transition might intervene and 
destroy the PS or HU correlations when $h,s=O(1)$. 

We are not able to rule out this possibility but several exact results indicate strongly that there is no such phase transition.
(i)~For $s\to-\infty$, the density correlations of the PASEP are known~\cite{popkov2011}: independently of the asymmetry $a_0$, there is a logarithmic effective interaction potential between particles which renders this state hyperuniform.  This implies that the jointly-biased SSEP is HU as $s\to-\infty$ (for all $h$).  (ii)~For $h=0$, a variational argument~\cite{garrahan2007} indicates that the SSEP phase separates for all $s>0$: the system can then access configurations where the total number of available hops remains finite as $L\to\infty$.  (iii)~Phase separation has been shown analytically for the totally asymmetric exclusion process ($a_0=1$): this transition corresponds to the appearance of ``shocks'' in response to a bias on the current~\cite{bodineau2005}.  For the SSEP, this establishes phase separation for all $B>0$ in the limit $h\to\infty$.  Combining these results establishes that the proposed phase diagram of Fig.~\ref{fig:phase}(a) is correct in all of the shaded regions: we cannot rule out other phase diagrams that are consistent with these constraints but this simple picture is the most likely scenario.  If the proposed Fig.~\ref{fig:phase}(a) is correct, the phase diagrams in Fig.~\ref{fig:phase}(b,c) follow from the exact mappings between biased SSEP and PASEPs.

\emph{Conclusion} -- %
Fig.~\ref{fig:phase} indicates that exclusion processes respond very strongly to biases $h$ and $s$, which almost always lead to either phase-separated or hyperuniform states.  The normal fluctuations that are familiar from equilibrium systems occur only under special high-symmetry conditions, such as $B=0$.  These results provide another example~\cite{zachary2011,chicken2014,levine-arxiv} of hyperuniformity emerging in non-equilibrium states, and they show that the dynamical phase transition identified in~\cite{appert2008} corresponds physically to the appearance of hyperuniformity.  More generally, the theory of fluctuating hydrodynamics  indicates that these dynamical phase transitions should be generic (``universal'') in systems with locally-conserved hydrodynamic variables such as energy or density.  The interplay between these phase transitions and the ``glass transitions" found previously in biased ensembles of trajectories~\cite{garrahan2007,garrahan2009,hedges2009} merits further study -- diffusive large-scale behaviour is not a necessary condition for those glass transitions~\cite{garrahan2007,elmatad2010}, but the analysis presented here indicates that phase-separated states may compete with homogeneous glassy states in systems that are biased to low activity.

We thank Fred van Wijland, Vivien Lecomte, and Juan P. Garrahan  for many useful discussions.  RLJ and IRT were supported by the EPSRC through grant EP/I003797/1.

\eject


\begin{appendix}

\section{Appendix A:\\ BROWNIAN HARD-PARTICLE MODEL}

In the Brownian hard-particle model, there are $N$ particles with positions $x_i(t)$, in a $1d$ box of size $L$, with periodic boundaries.  
Since the particles are hard, they do not overtake one another, so we assign particle indices such that $x_1(0) < x_2(0) < \dots < x_N(0)$.  According to the Langevin
dynamics, the centre of mass $\overline{x}(t) = \frac{1}{N} \sum_i x_i(t)$ undergoes free diffusion with diffusion constant $D_0/N$.  When evaluating the dynamical activity
$K$, we use $$\hat{x}_i(t) = x_i(t) - \overline{x}(t).$$ In cases where the particle travels around the periodic boundaries of the system, the position $x_i$ is defined so that it varies continuously, so we may
have $x_i<0$ or $x_i>L$; for the purposes of particle interactions, we use
the position of the particle within the box, which is $(x \hbox{ mod } L)$. 

 It is convenient to define $X_j = x_j - jl_0$ where
$l_0$ is the particle size.  The $X_j$ diffuse in a periodic box of size $L_0 = L-Nl_0$; in the equilibrium state, they are distributed independently and uniformly throughout the box.  
If particle indices are ignored, the trajectories of the $X_i$ are the same as those of independent freely diffusing particles (with no hard-core interaction).  This mapping is valid because in terms of the density field, a ``collision'' between two particles has the same effect as two particles diffusing past each other.  This means that multi-point space-time correlations of the density can in principle be calculated exactly.  However, the activity $K$ requires that we keep track of particle indices and collisions.  This means that the activity fluctuations in the model are not trivial, as is clear from Fig.~\ref{fig:fdm}.

\rlj{
When evaluating the structure
factor $S(q)$ for the \fdm\ we use the definition $\rho_q = \sum_j \ee^{-\ii q X_j}$ with $q=2m\pi/L_0$ and $m$ integer, so that $S(q) = \frac{1}{L_0} \langle |\rho_q|^2\rangle = N/L_0$ at equilibrium ($s=0$), independent of $q$.  We argue that this structure factor has the same small-$q$ behaviour as the structure factors calculated directly from the particle positions $x_i$, as follows.  The structure factor for small $q$ can be inferred from the fluctuations of the number of particles $\hat{n}_R$ within regions (subsystems) of size $R \sim 1/q$.  The `hat' notation
indicates that $\hat{n}_R$ is a fluctuating quantity.  For a homogeneous system, and assuming that $R$ is much larger than the particle spacing, we can equivalently consider the size $\hat{R}_n$ of a subsystem (or region) containing exactly $n$ particles.  If we take $n=\rho R$ then we expect
\begin{equation}
\frac{ \langle (\delta \hat{n}_R)^2 \rangle }{ \langle \hat{n}_R \rangle^2 } = \frac{ \langle (\delta \hat{R}_n)^2 \rangle }{ \langle \hat{R}_n \rangle^2 } .
\label{equ:fluct}
\end{equation}
At equilibrium, this relation is most easily proved via correlation-response formulae for the isothermal compressibility, but we argue here that it also holds in the out-of-equilibrium states found in biased ensembles.
To see this, define a local density $\hat{\rho}_R=\frac1R \int_0^{R} \rho(r) \mathrm{d}r$, which may either be written as $\hat{n}_R/R$ or obtained equivalently as $(n/\hat{R}_n)_{n=\rho R}$.  The variance
of the local density may then be written either in terms of $\hat{n}_R$ or $\hat{R}_n$.  We assume
(based on a self-averaging argument) that for large $n$ the fluctuations of $\hat{R}_n$ are small, from which we obtain (\ref{equ:fluct}).  Finally, we note that if $l_0$ is the size of the hard particles then
the probability distribution of $\hat{R}_n$ satisfies $P_{l_0}(\hat{R}_n) = P_{0}(\hat{R}_n-nl_0)$, where $P_{l_0}$ is the distribution for particles of size $l_0$ and $P_0$ is the distribution for point-particles.  Hence $\langle (\delta \hat{R}_n)^2 \rangle$ is equal in both representations, which is sufficient to establish that the structure factors $S(q)$ are equal (up to a multiplicative constant).
}%

To simulate the \fdm, we use a Monte Carlo (MC) scheme: in each MC move a particle $j$ is chosen at random and a displacement $\delta$ is chosen uniformly between $-\Delta$ and $\Delta$.  The move
is rejected if moving the particle to $x_j+\delta$ involves a particle overlap, otherwise it is accepted.  For small $\delta$, this scheme is equivalent to solving the Langevin dynamics of the interacting particles;
a time $\Delta t$ in the Langevin system corresponds to $\ell_0^2/(3\Delta^2)$ attempted MC moves per particle.  We take $\Delta=0.1l_0$.  
For the relatively high density ($\rho=0.88$) relevant for Fig.~1, it is convenient to use a continuous-time implementation of these dynamics in which all moves
are accepted~\cite{BKL}.

%

\section{Appendix B: FLUCTUATING HYDRODYNAMICS}

\subsection{Relation to microscopic models}

\begin{figure}
\includegraphics[width=8cm]{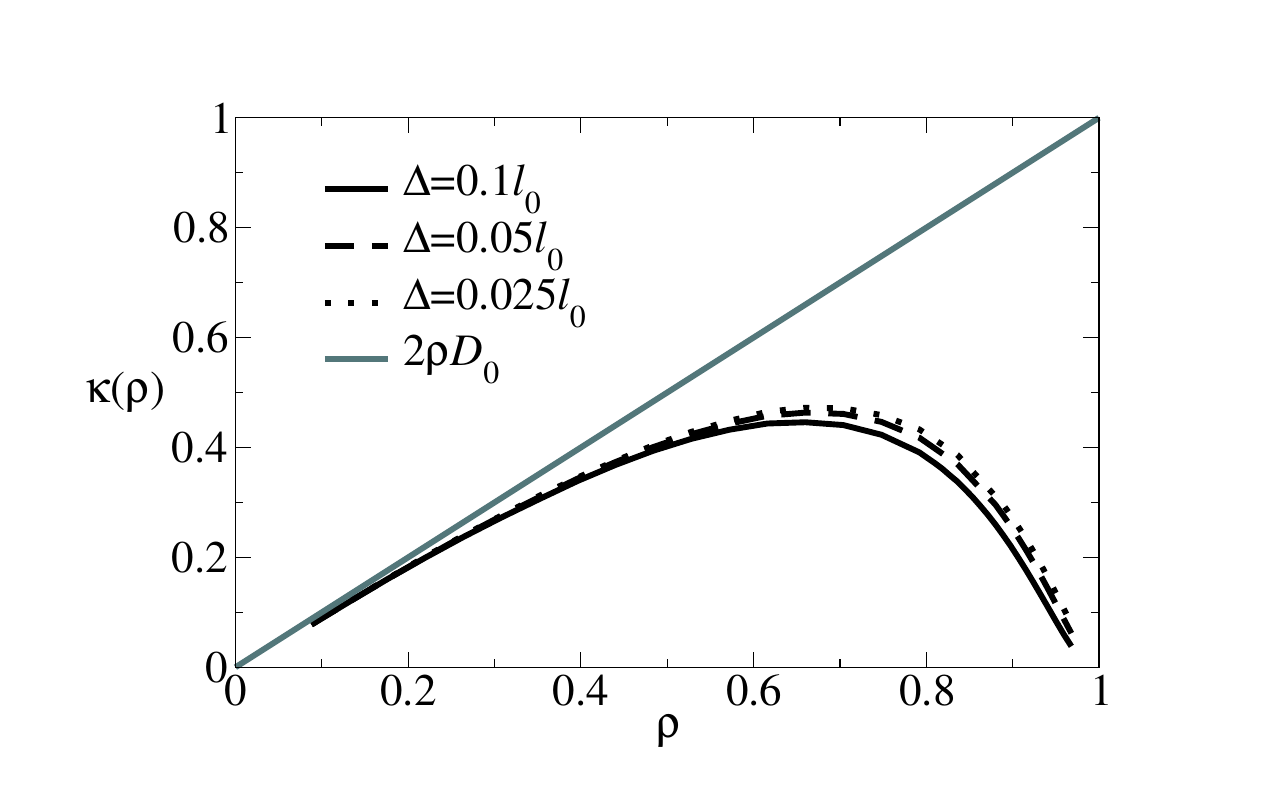}
\caption{Activity density $\kappa(\rho)$ for the \fdm, calculated numerically as $k(s=0)$ for different values of the 
maximal MC step $\Delta$.  The solid line shows the prediction for the dilute limit, for which $k(0)=2\rho D_0$.  [Note that the length and time units used in this work are $l_0=1$ and $\tau_0=1$, so 
that $D_0=1/2$.]  The non-monotonic shape
is similar to the behavior in the SSEP where $\kappa(\rho)=2\rho(1-\rho)$.  In the main text, we use an MC
step $\Delta = 0.1l_0$: this gives small quantitative deviations in the results from the limiting behaviour as $\Delta\to 0$, but the qualitative behavior is unchanged on reducing $\Delta$.
}\label{fig:k-rho}
\end{figure}

The relation between fluctuating hydrodynamic equation~(\ref{equ:fh-langevin}) and microscopic models has been discussed in several previous studies~\cite{bertini2001,tkl2008,appert2008,bertini-revs}.
The SSEP corresponds to fluctuating hydrodynamics with $D=1$, $\sigma=2\rho(1-\rho)$ and $\kkap=2\rho(1-\rho)$ (see for example~\cite{appert2008}, and note we 
have unit rates for both left and right hops).
For the PASEP, the fluctuating hydrodynamic theory applies only for the weakly-asymmetric model (sometimes called the WASEP), 
in which case $a=a_0=(r - \ell)/(r+\ell)$, and the theory applies only for very small $a=O(1/L)$. 

In the \fdm, the behaviour of the unbiased model corresponds to $\sigma=\rho$ and $D=1/2$; it is useful to compare with the small-$\rho$ limit of the SSEP, which 
reduces to the \fdm\ after a suitable rescaling of time.  
The function $\kkap(\rho)$ is not known analytically: we show numerical data in Fig.~\ref{fig:k-rho} from which we see that $\kkap''(\rho)<0$, 
as stated in the main text.

\subsection{Rescaled co-ordinates}

\newcommand{\rt}{\tilde{r}}
\newcommand{\tti}{\tilde{t}}
\newcommand{\qt}{\tilde{q}}
\newcommand{\ot}{\tilde{\omega}}

\newcommand{\vp}{\varphi}
\newcommand{\vphat}{\hat{\varphi}}

We take the hydrodynamic limit by rescaling lengths by a large (dimensionless) factor $R$, so our system of size $L$ maps
into a box of size $L/R$.  Typically one fixes $L/R$ and takes $L\to\infty$ but it will be convenient in the following to take first $L\to\infty$ and then later $R\to\infty$.
Let $\rt = r/R$ and $\tti=t/R^2$, and because $s$ has dimensions of a rate (inverse time) also $\lambda = s R^2$.  Let $\varphi(\rt,\tti) = \rho(\rt R, \tti R^2)$ and similarly $\vphat(\rt,\tti) = \rhohat(\rt R, \tti R^2)$.

Then the action in the path integral representation of the dynamics is
$\int\mathrm{d}r\mathrm{d}t\, {\cal L} = R^d \int\mathrm{d}\tilde{r}\mathrm{d}{\tilde t}\, \tilde{\cal L}$ with $\cal L$ given to quadratic order by (\ref{equ:lag-quad}), so that 
\begin{multline}
 \tilde{\cal L} \approx \ii \vphat (\partial_{\tti} - \nabla\cdot D_0\nabla)\delta\vp + \tfrac12 \sigma_0|\nabla\vphat|^2 +  \lambda k_0 
+ \tfrac12  \lambda \kkap''_0\delta\vp^2 ,
\label{equ:lag-phi}
\end{multline}
where the approximate equality emphasises that we have truncated at quadratic order.
\rlj{(Here $\nabla=\nabla_{\tilde r}$ acts on the rescaled co-ordinate: we omit the variable being used where this is unambiguous from the context.)}

Transforming to the Fourier domain, we write
$\vp_{\ot,\qt} = \int\mathrm{d}\rt\mathrm{d}\tti \, \delta\vp(\rt,\tti) \ee^{-\ii(\qt\cdot\rt - \ot\tti)}$. 
The inverse transform is
$\delta \vp(\tilde r) = \frac{R^{2+d}}{ L^d\tobs } \sum_{\ot}^\Omega \sum_{\qt}^{Q} \delta\vp_{\ot\qt} \, \ee^{\ii(\qt\cdot \rt-\ot\tti)}$.
The key point is that the theory is evaluated with fixed cutoffs $Q$ and $\Omega$ which are of order unity as $R\to\infty$.  This ensures (for example)
that loop corrections can always be neglected in perturbative calculations.  (The allowed wavectors are $\qt=2m\pi R/L$ for integer $m$ and similarly $\ot = 2n\pi R^2/\tobs$.)


\subsection{Activity bias}

To analyze the effect of the bias $s<0$, 
we start from 
(\ref{equ:lag-phi}).  We can write the action as
\begin{equation}
R^d \int\mathrm{d}\rt\mathrm{d}\tti \tilde{\cal L}  = \frac{R^{2d+2}}{L^d \tobs} \sum_{\ot,\qt} {\cal L}_{\ot,\qt} ,
\end{equation}
with (to quadratic order)
\begin{multline}
{\cal L}_{\ot,\qt}  = \ii\vphat_{\ot,\qt} (\ii\ot + D_0 \qt^2) \delta\vp_{-\ot,-\qt} + \tfrac12 \sigma_0\qt^2 |\vphat_{\ot,\qt}|^2 \\ +  \lambda \kkap_0 
+ \tfrac12  \lambda \kkap''_0|\delta\vp_{\ot,\qt}|^2 .
\label{equ:Lotqt}
\end{multline}

To calculate the structure factor, we must consider the integral $\int D\vp D\vphat |\vp_{\ot,\qt}|^2 \ee^{-R^d \int\mathrm{d}\rt\mathrm{d}\tti \tilde{\cal L} }$.  
\rlj{Making a saddle-point approximation, the behaviour depends only on the quadratic-order expansion (\ref{equ:Lotqt}),} %
and we obtain
\begin{equation}
\langle |\delta\vp_{\ot \qt}|^2 \rangle = \frac{L^d \tobs}{R^{2d+2}} \cdot \frac{ \sigma_0 \qt^2 }{ \ot^2 + (D_0 \qt^2)^2 + \lambda \qt^2 \sigma_0 \kkap''_0} .
\end{equation}

To obtain the equal-time fluctuations we define $\delta\vp_{\qt}(\tti)=\int\mathrm{d}\rt \delta\vp(\rt,\tti) \ee^{-\ii\qt\cdot\rt} = R^{-d} \delta\rho_q(t)$, with $q=\qt/R$.  We have
$\langle |\delta\vp_{\qt}(\tti)|^2 \rangle = \frac{R^4}{\tobs^2} \sum_{\ot} \langle |\delta\vp_{\ot \qt}|^2 \rangle$:
taking $\tobs\to\infty$ before any limit of large-$R$,
the sum over $\ot$ may be converted to an integral.  As long as the frequency cutoff $\Omega$ satisfies $\Omega^2 \gg (D_0 \qt^2)^2 + \lambda \qt^2 \sigma_0 \kkap''_0$, we obtain
\rlj{%
\begin{equation}
\langle |\delta\vp_{\qt}(t)|^2 \rangle = \frac{L^d}{R^{2d}} \cdot  \frac{ \sigma_0 \qt^2 }{ 2\sqrt{(D_0 \qt^2)^2 + \lambda \qt^2 \sigma_0 \kkap''_0}}.
\end{equation}
}%
\rlj{While the derivation of this  equation required a saddle-point approximation [equivalent to the truncation at quadratic order in (\ref{equ:lag-phi})], it can be shown that this approximation becomes exact in the limit of large-$R$.}
Finally, converting from the rescaled parameters $\vp$, $\qt$, $\lambda$ to the bare quantities $\rho$, $q$, $s$ yields (\ref{equ:fh-Sq}) of the main text.  The 
requirement that $R$ be large while $\qt$ and $\lambda$ are of order unity implies that (\ref{equ:fh-Sq}) is valid only for very small $q$ and $s$, as discussed in the main text.

\subsection{Current bias}

Now suppose that, instead of coupling $-s$ to activity, we couple $h$ to the particle current $J$.  Within fluctuating hydrodynamics, we write
$\partial_t \rho = -\nabla \cdot J$ with $J=-D\nabla\rho + \sqrt{\sigma}\eta$.  Using the method of Martin-Siggia-Rose-DeDominicis-Janssen~\cite{msr,tkl2008}, we arrive at a
path integral with Lagrangian
\begin{align}
{\cal L} 
&= \ii\rhohat(\partial_t - \nabla\cdot D\nabla)\rho - \tfrac{1}{2}\sigma ( \ii\nabla\rhohat + h )^2 + h D \nabla\rho  .
\label{equ:lag-h}
\end{align}
The system has periodic boundaries: if we assume (as expected)
that $D\nabla\rho = \nabla g$ for some $g=g(\rho)$, the last term in (\ref{equ:lag-h}) vanishes after integration over $r$.

Expanding about the homogeneous stationary profile as in the activity-biased case, the analogue of (\ref{equ:lag-quad}) of the main text is
\begin{multline}
{\cal  L }= \ii\rhohat(\partial_t - \nabla\cdot D_0\nabla)\delta\rho  - \tfrac{1}{2}\sigma_0 h^2 + \tfrac{1}{2}\sigma_0(\nabla\rhohat)^2 \\
 - \ii\sigma'_0\delta\rho (h\cdot\nabla\rhohat) - \tfrac14 h^2\sigma_0''\delta\rho^2  .
\end{multline}
Stability of the homogeneous profile requires $\sigma''_0\leq0$, which is the case for exclusion processes and the \fdm.
To proceed, we rescale co-ordinates as in the previous section, defining in addition $\tilde{h}=hR$.   The calculation is almost identical so
we give only a brief discussion: we find
\rlj{
\begin{equation}
\langle |\delta\vp_{\ot \qt}|^2 \rangle = \frac{L^d \tobs}{R^{2d+2}} \cdot \frac{ \sigma_0 \qt^2 }{ (\ot - \sigma_0'\tilde{h}\cdot\qt)^2 + (D_0 \qt^2)^2 +  \tfrac12 \tilde{h}^2 \qt^2 \sigma_0\sigma_0''} .
\label{equ:phiphi-h}
\end{equation}
}
To obtain the structure factor, we perform a frequency integral, and the term $\tilde{h}\sigma_0'\qt$ plays no role since it
can be absorbed by a shift of the integration variable.  Hence we arrive at the result for the structure factor (in terms of the bare variables):
\rlj{%
\begin{equation}
S(q) = \frac{ \sigma_0 q^2 }{ 2\sqrt{(D_0 q^2)^2 - \tfrac12 h^2q^2 \sigma_0  \sigma''_0}} .
\end{equation}
}%
(Note $\sigma''<0$, by assumption.)
This is the same form as (\ref{equ:fh-Sq}) of the main text,  but with
$-s\to h^2/2$ and $\kkap_0'' \to \sigma_0''$.  
Hence, as long as $\sigma_0'',\kkap_0''<0$, the effect of the current bias $h$ is the same as the bias to higher-than-average activity, $s<0$.

\subsection{Joint bias} 

The analysis for a joint bias on activity and current is a trivial extension of the previous cases.
Assuming a homogeneous state, we obtain 
\rlj{
\begin{equation}
\langle |\delta\vp_{\ot \qt}|^2 \rangle =  \frac{L^d \tobs}{R^{2d+2}} \frac{ \sigma_0 \qt^2 }{ (\ot - \sigma_0' h\cdot \qt)^2 + (D_0 \qt^2)^2 + B\qt^2},
\end{equation}
with $B= s\kkap_0''\sigma_0 - \sigma''_0 h^2 \sigma_0/2$.  The self-consistency condition for homogeneity is $B>0$.  
The similarity with (\ref{equ:phiphi-h}) allows straightforward calculation of the structure factor in these ensembles.}

\subsection{Scaling of $k(s)$ for $s\leq0$}

We now show how the activity $k(s)$ can be calculated within fluctuating hydrodynamics.  The relevant expression is given by expanding $k(s) = \langle \kkap(\rho) \rangle_s$ to quadratic
order in $\delta\rho$:
\begin{equation}
k(s) = \kkap_0 + \frac12  \kkap_0'' \langle \delta\rho(r,t)^2  \rangle_s + \dots
\end{equation}
Hence, working at quadratic order
\begin{equation}
k(s) - k(s=0) =  \kkap_0'' \frac{1}{2L^d}   \sum_q [S(q) - S_{s=0}(q)] .
\end{equation}
We define $\Delta k = k(s) - k(s=0)$: in terms of the rescaled hydrodynamic variables, we obtain
\begin{equation}
\Delta k =  \kkap_0''  \frac{R^{2d}}{2L^{2d}}\sum_{\qt} \left[ \langle |\delta\vp_{\qt}(t)|^2 \rangle_s - \langle |\delta\vp_{\qt}(t)|^2 \rangle_0 \right] .
\end{equation}
Taking $L\to\infty$ at fixed $R$, we can convert the sum to an integral arriving at
\rlj{
\begin{equation}
\Delta k =  \frac{\kkap_0''\sigma_0}{4D_0} \, \frac{1}{(2\pi R)^d} \int_{|\qt|<Q}\!\!\mathrm{d}\qt \left[ \frac{\qt}{\sqrt{\qt^2+(\lambda\sigma_0\kkap_0''/D_0^2)}} -1 \right] .
\label{equ:DKx}
\end{equation}
}%
It is convenient to define $x = \lambda\sigma_0\kkap_0''/D_0^2$, so $x \propto \lambda$.

For $d=1$ we obtain 
\rlj{
\begin{equation}
\Delta k =  \frac{\kkap_0''\sigma_0}{4\pi D_0 R} \left( -\sqrt{x} + \sqrt{Q^2+x} - Q \right).
\end{equation}
}%
The leading $\sqrt{x}$ term is non-analytic at $x=0$ (which corresponds to $\lambda=0$): in terms
of the bare parameters, this gives the $\Delta k(s) \propto \sqrt{-s}$ result quoted in the main text.

For $d>1$, the integral gives a singular behaviour proportional to $x^{d/2}$ in odd dimensions, and $x^{d/2}\log x$ in even dimensions.  There are also analytic ``non-universal" ($Q$-dependent) terms 
that lead to a polynomial dependence on $x$.  For example in $d=3$
\begin{equation}
\Delta k \propto \tfrac23 x^{3/2} + \tfrac13 \left[  (Q^2  - 2x)\sqrt{ Q^2 + x} - Q^3 \right] .
\end{equation}
where the leading behaviour at small $x$ is a non-universal term at $O(x)$ but the first singular contribution is the universal $x^{3/2}$ contribution from small wavevectors.
Similarly for $d=2$,
\begin{equation}
\Delta k \propto -\tfrac14 x \log x + \tfrac12 \left[ x\log( Q + \sqrt{Q^2+x} ) - Q\sqrt{Q^2+x} \right] .
\end{equation}
The behavior in higher dimensions can be obtained analogously by repeated integration by parts, starting from (\ref{equ:DKx}).

\section{Appendix C:\\ EXACT MAPPINGS FOR EXCLUSION PROCESSES}

As in the main text the jointly biased SSEP (with periodic boundaries) is associated with an operator~\cite{spohn1999,lecomte2005,appert2008}
\begin{multline}
W_{\rm S}(s,h) = \sum_i \ee^{h-s}\sig^-_{i}\sig^+_{i+1} + \ee^{-h-s}\sig^-_{i}\sig^+_{i-1}  \\  - 2 n_i (1- n_{i+1} ) .
\end{multline}
Similarly, the relevant operator for the current-bisaed PASEP is
\begin{multline}
W_{\rm AJ}(\ell,r,\tilde h)=\sum_i r\ee^{\tilde h}\sig^-_{i}\sig^+_{i+1} + \ell\ee^{-\tilde h}\sig^-_{i}\sig^+_{i-1} \\ + (\ell+r) (n_i n_{i+1} - N) .
\end{multline}
Note the Gallavotti-Cohen symmetry~\cite{gallavotti-cohen,spohn1999}: setting $\ee^{\tilde h}=\ell/r$ recovers the original unbiased model but with the opposite bias.

For correspondence with the jointly biased SSEP we require $r\ell=\ee^{-2s}$ and $r/\ell=\ee^{2(h-\tilde h)}$, as well as $r+\ell=2$.  Hence $r=\ee^{h-\tilde h-s}$
and $\ell=\ee^{-h+\tilde h-s}$.  Adding gives $\cosh(h-\tilde h) = \ee^{s}$: the case $\cosh(h)=\ee^s$ reduces to an unbiased PASEP.  For small $h$, this means $s\approx h^2/2$
which corresponds to the condition $B=0$ in analysis of fluctuating hydrodynamics.
\rlj{The general mapping from SSEP to current-biased PASEP requires $s>0$, 
and  there are typically two solutions for $\tilde h$:}
each point in the right half plane of Fig.~\ref{fig:phase}(a) therefore maps to two points in Fig.~\ref{fig:phase}(b), one with $a_0= (r-\ell)/(r+\ell)>0$
and the other with $a_0<0$. 

If we instead consider an activity-biased PASEP, the relevant operator is
\begin{multline}
W_{\rm AK}(\ell,r,\tilde s)=\sum_i r\ee^{-\tilde s}\sig^-_{i}\sig^+_{i+1} + \ell\ee^{-\tilde s}\sig^-_{i}\sig^+_{i-1}  \\ - (\ell+r) n_i (1 -n_{i+1} ) .
\end{multline}
For correspondence with $W_{\rm S}$ we require $r/\ell=\ee^{2h}$ and $r\ell=\ee^{2(\tilde s-s)}$ and again $r+\ell=2$.  Taking the first and third of these gives
the asymmetry parameter as $a_0  = \tanh(h)$: the current bias on the SSEP sets the asymmetry of the PASEP.
On the other hand, taking the first and second constraints, $r=\ee^{h-s+\tilde s}$ and $\ell=\ee^{-h-s+\tilde s}$ and given $r+\ell=2$ we have $\ee^{-\tilde s} = \ee^{-s} \cosh(h)$.
As before, we recover an unbiased PASEP if $\ee^{-s}\cosh(h)=1$.

The general case of a PASEP with a joint bias on current and activity is a simple generalisation.   
Given that we have fixed the time unit in the SSEP so that the coefficient of the diagonal term $n_i (1-n_{i+1})$ is always $2$,
the jointly biased SSEP is a two-parameter family of models dependent on $(h,s)$.  
Fixing the time unit in the PASEP in the same way, the jointly biased PASEP has $3$ parameters: $(a_0,\tilde{h},\tilde{s})$.  However, the only free parameters for the 
relevant operator are the coefficients of the $\sigma_i^-\sigma_{i+1}^+$ and $\sigma_i^+\sigma_{i+1}^-$ terms.  
Hence every jointly biased SSEP can be mapped into a one-parameter family of PASEPs.

We note that these mappings also hold in $d>1$ if the activity bias $s$ couples to the total number of hops along just one Cartesian direction.

%
%

\subsection{Weak asymmetry and fluctuating hydrodynamics}

The mapping between jointly biased SSEP and the unbiased weakly-asymmetric exclusion process (WASEP) can also be accomplished at the fluctuating hydrodynamic level.  
E.g.\ for a WASEP with asymmetry-parameter $a$ the Lagrangian is
\begin{equation}
{\cal L}_{\rm WA} = \ii\rhohat(\partial_t - \nabla\cdot D\nabla)\rho + \ii\rhohat a\cdot\nabla\sigma +  \tfrac12 \sigma (\nabla\rhohat)^2  .
\end{equation}
Integrating by parts on the term with just one gradient and then completing the square yields
\begin{equation}
{\cal L} _{\rm WA}= \ii\rhohat(\partial_t - \nabla\cdot D\nabla)\rho + \tfrac12 \sigma(\ii\nabla\rhohat-a)^2 + \
\tfrac12 a^2\sigma.
\end{equation}
This a jointly biased SSEP with a current-bias $h=a$ and an activity bias $s=a^2/2$ [we have $k(\rho)=\sigma(\rho)$] so that $s=h^2/2$. This is consistent with the fact that any unbiased WASEP has normal fluctuations, so must lie on the line $B=0$.

\end{appendix}


\begin{thebibliography}{99}

\bibitem{spohn83}
H. Spohn, J. Phys. A {\bf 16}, 4275 (1983)

\bibitem{btw}
P.~Bak, C.~Tang and K.~Wiesenfeld, Phys. Rev. Lett. {\bf 59}, 381 (1987).

\bibitem{bodineau2004}
T.~Bodineau and B.~Derrida, Phys. Rev. Lett. {\bf 92}, 180601 (2004).

\bibitem{garrahan2007}
J.~P.~Garrahan, R.~L.~Jack, V.~Lecomte, E.~Pitard, K.~van~Duijvendijk and F.~van~Wijland,
Phys. Rev. Lett. {\bf 98}, 195702 (2007)

\bibitem{lecomte2005}
V.~Lecomte, C.~Appert-Roland and F.~van~Wijland, Phys. Rev. Lett. {\bf 95}, 010601 (2005)

\bibitem{bodineau2008}
T.~Bodineau, B.~Derrida, V.~Lecomte and F.~van~Wijland, J. Stat. Phys. {\bf 133},  1013 (2008).

\bibitem{hedges2009}
L.~O.~Hedges, R.~L.~Jack, J.~P.~Garrahan and D.~Chandler, 
Science {\bf 323}, 1309 (2009).

\bibitem{bertini2001}
L. Bertini, A. De Sole, D. Gabrielli, G. Jona-Lasinio, and C. Landim,
Phys. Rev. Lett. {\bf 87}, 040501 (2001); J. Stat. Phys. {\bf 107}, 625 (2002).

\bibitem{bertini2005}
L. Bertini, A. De Sole, D. Gabrielli, G. Jona-Lasinio, and C. Landim,
Phys. Rev. Lett. {\bf 94}, 030601 (2005).

\bibitem{tkl2007}
J. Tailleur, J. Kurchan and V. Lecomte, Phys. Rev. Lett. {\bf 99}, 150602 (2007)

\bibitem{hurtado2011-pnas}
P.~I.~Hurtado, C.~Perez-Espigares, J.~J.~del Pozo and P.~L.~Garrido, PNAS {\bf 108}, 7704 (2011)

\bibitem{bertini-revs}
L. Bertini, A.~De Sole, D.~Gabrielli, G.~Jona-Lasinio and C.~Landim,
J. Stat. Phys. {\bf 135}, 857 (2009);
arXiv:1404.6466 (2014)

\bibitem{torquato2003}
S.~Torquato and F.~H.~Stillinger,
Phys. Rev, E {\bf 68}, 041113 (2003).

\bibitem{florescu2009}
M.~Florescu, S.~Torquato and P.~J.~Steinhardt, PNAS {\bf 106}, 20658 (2009)

\bibitem{zachary2011}
C.~E.~Zachary, Y.~Jiao and S.~Torquato, Phys. Rev. Lett. {\bf 106}, 178001 (2011)

\bibitem{berthier2011}
L.~Berthier, P.~Chaudhuri, C.~Coulais, O.~Dauchot and P.~Sollich,
Phys. Rev. Lett. {\bf 106}, 120601 (2011)

\bibitem{man2013}
W.~Man, M.~Florescu, K.~Matsuyama, P.~Yadak, G.~Nahal, S.~Hashemizad,
E.~Williamson, P.~Steinhardt, S.~Torquato and P.~Chaikin,
Optics Express {\bf 21}, 19972 (2013).

\bibitem{chicken2014}
Y.~Jiao, T.~Lau, H.~Hatzikiriou, M.~Meyer-Hermann, J.~C.~Corbo and S.~Torquato,
Phys. Rev. E {\bf89}, 022721 (2014)

\bibitem{levine-arxiv}
D.~Hexner and D.~Levine, arXiv:1407.0146.

\bibitem{tkl2008}
J. Tailleur, J. Kurchan and V. Lecomte, J. Phys. A {\bf 41}, 505001 (2008)

\bibitem{appert2008}
C.~Appert-Rolland, B.~Derrida, V. Lecomte and F.~van Wijland,
Phys. Rev. E {\bf 78}, 021122 (2008)

\bibitem{spohn1999}
J.~L.~Lebowitz and H.~Spohn, J. Stat. Phys. {\bf 95}, 333 (1999).

\bibitem{touchette2013}
R.~Ch\'etrite and H.~Touchette, arXiv:1405.5157.

\bibitem{tps}
P.~G.~Bolhuis, D.~Chandler, C.~Dellago and P.~L.~Geissler,
Ann. Rev. Phys. Chem. {\bf53}, 291 (2002).

\bibitem{lecomte2012}
V.~Lecomte, J.~P.~Garrahan and F.~van~Wijland, J. Phys. A {\bf 45}, 175001 (2012).


\bibitem{hurtado-rev}
P.~I.~Hurtado, C.~P.~Espigares, J.~J.~del Pozo, P.~L.~Garrido,
J. Stat. Phys. {\bf 154}, 214 (2014)

\bibitem{eyink1990}
G.~L.~Eyink, J. Stat. Phys. {\bf 61}, 533 (1990).

\bibitem{kmp}
C. Kipnis, C. Marchioro, E. Presutti, 
J. Stat. Phys. {\bf 27}, 65 (1982).

\bibitem{lecomte2010}
V. Lecomte, A. Imparato, and F. van Wijland, Prog Theor Phys. Supp. {\bf 184}, 276 (2010).

\bibitem{propensity}
A.~Widmer-Cooper, P.~Harrowell and H.~Fynewever, Phys, Rev. Lett. {\bf93} 135701 (2004).

\bibitem{garrahan2009}
J.~P.~Garrahan, R.~L.~Jack, V.~Lecomte, E.~Pitard, K.~van Duijvendijk, and F.~van Wijland,
J. Phys. A {\bf 42}, 075007 (2009).

\bibitem{jack2014-east}
R.~L.~Jack and P.~Sollich, J. Phys. A {\bf 47}, 015003 (2014).

\bibitem{jack2010rom}
R.~L.~Jack and J.~P.~Garrahan, Phys. Rev. E {\bf 81}, 011111 (2010).

\bibitem{stinch-review}
R.~Stinchcombe, Adv. Phys. {\bf 50}, 431 (2001).

\bibitem{gallavotti-cohen}
G.~Gallavotti and E.~G.~D.~Cohen, J. Stat. Phys. {\bf 80}, 931 (1995).

\bibitem{popkov2011}
V.~Popkov and G.~M.~Sch\"utz, J. Stat. Phys. {\bf 142}, 627 (2011).

\bibitem{bodineau2005}
T.~Bodineau and B.~Derrida, Phys. Rev. E {\bf 72}, 066110 (2005).

\bibitem{elmatad2010}
Y.~S.~Elmatad, R.~L.~Jack, J.~P.~Garrahan and D.~Chandler,
PNAS {\bf 107}, 12793 (2010).

\bibitem{msr}
P.~C.~Martin, E.~D.~Siggia and H.~A.~Rose, Phys. Rev. A {\bf 8}, 423 (1973);
C. De Dominicis, Lett. Nuovo Cimento, {\bf 12}, 567 (1975);
H.-K.~Janssen, Z. Phys. B {\bf 23}, 377 (1976).

\bibitem{BKL}
A.~B.~Bortz, M.~H.~Kalos and J.~L.~Lebowitz, J. Comp. Phys. {\bf 17}, 10 (1975);
M.~E.~J.~Newman and G.~T.~Barkema, \emph{Monte Carlo Methods in Statistical Physics}, (OUP, Oxford, 1999).


\end{thebibliography}
\end{document}